\documentstyle[12pt]{article}
\newread\epsffilein    
\newif\ifepsffileok    
\newif\ifepsfbbfound   
\newif\ifepsfverbose   
\newdimen\epsfxsize    
\newdimen\epsfysize    
\newdimen\epsftsize    
\newdimen\epsfrsize    
\newdimen\epsftmp      
\newdimen\pspoints     
\def\epsfbox#1{\global\def\epsfllx{72}\global\def\epsflly{72}%
   \global\def\epsfurx{540}\global\def\epsfury{720}%
   \def\lbracket{[}\def\testit{#1}\ifx\testit\lbracket
   \let\next=\epsfgetlitbb\else\let\next=\epsfnormal\fi\next{#1}}%
\def\epsfgetlitbb#1#2 #3 #4 #5]#6{\epsfgrab #2 #3 #4 #5 .\\%
   \epsfsetgraph{#6}}%
\def\epsfnormal#1{\epsfgetbb{#1}\epsfsetgraph{#1}}%
\def\epsfgetbb#1{%
\openin\epsffilein=#1
\ifeof\epsffilein\errmessage{I couldn't open #1, will ignore it}\else
   {\epsffileoktrue \chardef\other=12
    \def\do##1{\catcode`##1=\other}\dospecials \catcode`\ =10
    \loop
       \read\epsffilein to \epsffileline
       \ifeof\epsffilein\epsffileokfalse\else
          \expandafter\epsfaux\epsffileline:. \\%
       \fi
   \ifepsffileok\repeat
   \ifepsfbbfound\else
 \ifepsfverbose\message{No bounding box comment in #1; using defaults}\fi\fi
   }\closein\epsffilein\fi}%
\def\epsfclipstring{}
\def\epsfsetgraph#1{%
   \epsfrsize=\epsfury\pspoints
   \advance\epsfrsize by-\epsflly\pspoints
   \epsftsize=\epsfurx\pspoints
   \advance\epsftsize by-\epsfllx\pspoints
   \epsfxsize\epsfsize\epsftsize\epsfrsize
   \ifnum\epsfxsize=0 \ifnum\epsfysize=0
      \epsfxsize=\epsftsize \epsfysize=\epsfrsize
      \epsfrsize=0pt
     \else\epsftmp=\epsftsize \divide\epsftmp\epsfrsize
       \epsfxsize=\epsfysize \multiply\epsfxsize\epsftmp
       \multiply\epsftmp\epsfrsize \advance\epsftsize-\epsftmp
       \epsftmp=\epsfysize
       \loop \advance\epsftsize\epsftsize \divide\epsftmp 2
       \ifnum\epsftmp>0
          \ifnum\epsftsize<\epsfrsize\else
             \advance\epsftsize-\epsfrsize \advance\epsfxsize\epsftmp \fi
       \repeat
       \epsfrsize=0pt
     \fi
   \else \ifnum\epsfysize=0
     \epsftmp=\epsfrsize \divide\epsftmp\epsftsize
     \epsfysize=\epsfxsize \multiply\epsfysize\epsftmp
     \multiply\epsftmp\epsftsize \advance\epsfrsize-\epsftmp
     \epsftmp=\epsfxsize
     \loop \advance\epsfrsize\epsfrsize \divide\epsftmp 2
     \ifnum\epsftmp>0
        \ifnum\epsfrsize<\epsftsize\else
           \advance\epsfrsize-\epsftsize \advance\epsfysize\epsftmp \fi
     \repeat
     \epsfrsize=0pt
    \else
     \epsfrsize=\epsfysize
    \fi
   \fi
   \ifepsfverbose\message{#1: width=\the\epsfxsize, height=\the\epsfysize}\fi
   \epsftmp=10\epsfxsize \divide\epsftmp\pspoints
   \vbox to\epsfysize{\vfil\hbox to\epsfxsize{%
      \ifnum\epsfrsize=0\relax
        \includegraphics{#1}%
      \else
        \epsfrsize=10\epsfysize \divide\epsfrsize\pspoints
        \includegraphics{#1}%
      \fi
      \hfil}}%
\global\epsfxsize=0pt\global\epsfysize=0pt}%
{\catcode`\%=12 \global\let\epsfpercent=
\long\def\epsfaux#1#2:#3\\{\ifx#1\epsfpercent
   \def\testit{#2}\ifx\testit\epsfbblit
      \epsfgrab #3 . . . \\%
      \epsffileokfalse
      \global\epsfbbfoundtrue
   \fi\else\ifx#1\par\else\epsffileokfalse\fi\fi}%
\def\epsfempty{}%
\def\epsfgrab #1 #2 #3 #4 #5\\{%
\global\def\epsfllx{#1}\ifx\epsfllx\epsfempty
      \epsfgrab #2 #3 #4 #5 .\\\else
   \global\def\epsflly{#2}%
   \global\def\epsfurx{#3}\global\def\epsfury{#4}\fi}%
\def\epsfsize#1#2{\epsfxsize}
\newcommand{\lwig}{\mbox{\,\raisebox{.3ex}
{$<$}$\!\!\!\!\!$\raisebox{-.9ex}{$\sim$}\,}}
\newcommand{\gwig}{\mbox{\,\raisebox{.3ex}
{$>$}$\!\!\!\!\!$\raisebox{-.9ex}{$\sim$}}\,}

\font\tenrm=cmr10
\font\tenit=cmti10
\font\elevenbf=cmbx10 scaled\magstep 1
\font\elevenrm=cmr10 scaled\magstep 1
\font\elevenit=cmti10 scaled\magstep 1

\font\ninerm=cmr9

\font\sevenrm=cmr7
\font\fiverm=cmr5

\textwidth 6.0in
\textheight 8.3in
\pagestyle{plain}
\topmargin -0.25truein
\oddsidemargin 0.30truein
\evensidemargin 0.30truein
\raggedbottom

\renewenvironment{thebibliography}[1]
{ \elevenrm
   \begin{list}{\arabic{enumi}.}
    {\usecounter{enumi} \setlength{\parsep}{0pt}
     \setlength{\itemsep}{3pt} \settowidth{\labelwidth}{#1.}
     \sloppy
    }}{\end{list}}

\parindent=3pc
\baselineskip=10pt
\newcommand{\capfont}{\baselineskip=12pt \tenrm
\textfont0=\tenrm \scriptfont0=\sevenrm \scriptscriptfont0=\fiverm }

\begin{document}

\begin{titlepage}
\vspace{0.75cm}

\begin{flushleft} DESY 94-026
\qquad \qquad\qquad
\qquad\qquad \qquad\qquad
\qquad\qquad \qquad \quad ~~~\qquad
ISSN 0418-9833 \\ UCLA/94/TEP/9 \\
March 1994
\end{flushleft}
\begin{center}
\vglue 2.0cm
{
 {\elevenbf        \vglue 10pt
    ASTROPHYSICAL SEARCHES FOR EXOTIC  \\
               \vglue 3pt
    PHENOMENA IN ULTRAHIGH ENERGY \\
               \vglue 3pt
    NEUTRINO--NUCLEON SCATTERING\footnote{
\ninerm
\baselineskip=11pt
Invited talk presented by A. Ringwald at {\it
3rd NESTOR International Conference}, 19-21 October 1993,
Pylos, Greece. To appear in the proceedings of the conference.}
\\}\vspace{0.75cm}

\vglue 1.0cm
{\tenrm D.A. MORRIS\footnote{email: morris@madonna.physics.ucla.edu}\\}
\baselineskip=13pt
{\tenit University of California, Los Angeles, 405 Hilgard Ave.\\}
\baselineskip=12pt
{\tenit Los Angeles, CA 90024, U.S.A.\\}
\vglue 0.3cm
{\tenrm and\\}
\vglue 0.3 cm
{\tenrm A. RINGWALD\footnote{email: ringwald@HP-Cluster.desy.de}\\}
\baselineskip=13pt
{\tenit DESY, Notkestrasse 85\\}
\baselineskip=12pt
{\tenit D--22603 Hamburg, Germany\\}}
\vglue 0.8cm
{\tenrm ABSTRACT}

\end{center}

\vglue 0.3cm
{\rightskip=3pc
 \leftskip=3pc
 \tenrm\baselineskip=12pt
 \noindent
We investigate the potential of near--future neutrino
telescopes like NESTOR for searches for exotic processes in
ultrahigh energy neutrino--quark scattering.
We consider signatures such as muon bundles and/or contained
cascades from the nonperturbative production
of multiple weak gauge bosons in the Standard Model,
compositeness and leptoquark production.
}
\end{titlepage}

{\elevenbf\noindent 1. Introduction}
\vglue 0.4cm
\baselineskip=18pt
\elevenrm
\setcounter{page}{2}
The introduction of neutrino telescopes such as AMANDA$^1$,
Baikal NT--200$^2$, DU\-MAND$^3$, and NESTOR$^4$ will
begin a new era in astronomy. In this talk we
report on how these detectors might also enlarge
our knowledge of particle physics. In particular,
we wish to explore the sensitivity of NESTOR to exotic
processes in ultrahigh energy neutrino--quark scattering  at
center of mass energies in the multi--TeV region.
Such energies will only become accessible to terrestrial
accelerators after the commissioning of CERN's proposed
Large Hadron Collider (LHC).

Compared to the sophistication of compact, accelerator--based
detectors, water Cherenkov detectors such as NESTOR concentrate
on relatively specific characteristics of individual events.
To compensate for this limitation, we will
restrict our discussion to three examples of
exotic phenomena with particularly
striking signatures: 1) the multiple production of weak
gauge bosons (W,Z) due
to nonperturbative features of the standard electroweak theory
(multi-W processes),
ii) multiple production of quarks
and leptons
due to compositeness of quarks and leptons, and
iii) the formation and decay of
leptoquarks in the neutrino--quark subprocess.

To be quantitative, we will adopt a phenomenological approach to
investigate the existence of exotic processes. We will parametrize the
anticipated features of new phenomena and then ask what region
of parameter space is accessible to near--future experiments. Our
presentation will proceed as follows. In section 2 we review the
phenomenology expected if nonperturbative multiple weak gauge boson
production exists. Since this phenomena exists entirely within
the framework of the Standard Model, we devote considerable attention
to it. As our discussion of multi--W phenomena progresses, we introduce
neutrino fluxes and describe the geometry of the stage--2 NESTOR array
which we  use in our subsequent calculations. In sections 3 and 4
we discuss compositeness and leptoquarks respectively.

\vglue 0.6cm
{\elevenbf\noindent 2.
Neutrino--Initiated Multi--W(Z) Production}
 \vglue 0.4cm
{\elevenit\noindent 2.1.  Nonperturbative
 Multi--W(Z) Production}
\vglue 0.4cm
\baselineskip=18pt
\elevenrm
A few years ago it was realized$^{5-8}$
that, even within the context of
a weakly coupled Standard Model,
leading--order perturbative calculations of the
inelastic scattering of quarks and leptons
involving the production of $\gwig~{\cal O}(\alpha_W^{-1}) \simeq 30$
weak gauge bosons result in an explosive (and unitarity violating)
growth of the associated parton-parton cross section above
center of mass energies
$ \gwig~ {\cal O}(\alpha_W^{-1} M_W ) \simeq 2.4$~TeV.
It was found that, to leading--order in $\alpha_W,$
amplitudes for the production
of $n_W$ W and Z bosons ({\it e.g.}, in neutrino--quark collisions)
exhibit a factorial growth in $n_W$.
This happens both for baryon and lepton number (B+L)
conserving amplitudes$^{7-9}$,

\begin{equation}
\label{eq:bcons}
 {\cal A}_{\rm lead.\ ord.}^{\rm B+L\  cons.}(\nu + q\to
\ell +q+
n_W\ W(Z) )
\propto
n_W!\ \alpha_W^{n_W/2}\ \Biggl(
{n_W\over \sqrt{\hat s}}\Biggr)^{n_W} ,
\end{equation}
as well as for anomalous B+L violating amplitudes$^{5,6}$,
\begin{eqnarray}
\label{eq:bviol}
\lefteqn{ {\cal A}_{\rm lead.\ ord.}
^{\rm B+L\ viol. }
(\nu + q\to 8\overline{q} + 2\overline{\ell } +
n_W\ W(Z) )
\propto }\nonumber \\
 & & \qquad \qquad \qquad
\qquad \qquad \qquad n_W!\ \alpha_W^{n_W/2}\ \Biggl(
{\sqrt{\hat s}\over n_Wm_W}\Biggr)^{n_W}
{{\rm e}^{-2\pi /\alpha_W}\over m_W^{n_W}} \, \, .\nonumber \\ & &
\end{eqnarray}

Perturbation theory
applied to large order processes ({\it e.g.}, like the production of
${\cal O}(\alpha_W^{-1})$ weak bosons) breaks down somewhere
in the multi--TeV range.
It is presently an open theoretical
question whether large--order weak interactions
become strong at this energy scale
(in the sense of having observable cross sections)
or whether they remain unobservably small at
all energies. The answer almost certainly lies beyond the realm
of conventional perturbative techniques
(see Ref.~10 for an overview).
A quantitative
consideration of experimental constraints on
multi--W production is clearly desirable.
In this section we
report on our recent investigations of this question$^{11,12}$
(see also Refs. 13--15).

\vglue 0.6cm
{\elevenit\noindent 2.2. Parametrization of Multi--W(Z) Production}
\vglue 0.4cm
In the absence of a reliable first--principles calculation
of multi--W production, we
parametrize
the quark--quark or neutrino--quark
cross section for multi--W production by
\begin{equation}
\label{eq:workhyp}
\hat \sigma_{\rm multi-W} = \hat \sigma_0\
\Theta
\bigl( \sqrt{\hat s} -\sqrt{\hat s_0 } \bigr) .
\end{equation}
This two--parameter working hypothesis frees us from specifying an
underlying (and most likely nonperturbative) mechanism for
multi--W production. According to our parameterization
the parton-parton cross section for multi-W production turns
on with a strength $\hat{\sigma}_0$ above a parton-parton
center of mass threshold energy $\sqrt{\hat{s}_0}.$
By convoluting the subprocess cross section of Eq.~(\ref{eq:workhyp})
with the appropriate quark distribution functions, one obtains the
multi--W production cross sections appropriate for nucleons.
For definiteness, we will assume
throughout this talk that $\hat\sigma_0$ refers to the production
of exactly 30 W bosons; allowing for the production of variable numbers
of W's (and Z's and possibly prompt photons)
is straightforward but is an unnecessary complication
at the level of our investigation.

An optimistic range of parameters to consider might encompass
\begin{equation}
\begin{array}{lcccl}
\label{eq:threshold}
\displaystyle{m_W \over \alpha_W } \simeq 2.4~\mbox{TeV} &
\leq &
\sqrt{\hat s_0} &
\leq &
40~\mbox{TeV}, \\
 & & & & \\
\label{eq:partoncro}
\displaystyle{\alpha_W^2 \over m_W^2 } \simeq 100 ~\mbox{pb} &
\leq &
\hat \sigma_0  &
\leq &
\sigma^{pp}_{\rm inel} \times
\left( \displaystyle{1~\mbox{GeV}\over m_W}\right)^2 \simeq
10~\mu\mbox{b}.
\end{array}
\end{equation}
The lower limit of $\sqrt{\hat s_0}$ is suggested by the
energy scale at which perturbation theory becomes unreliable$^{7-9}$
whereas the upper range is of the order of
the sphaleron mass$^{16}$, the characteristic scale of
anomalous electroweak B+L violation.
Unitarity arguments suggest$^{17}$
that the cross section for
nonperturbative multi--W(Z) production
is always exponentially suppressed, ${\hat\sigma}_0\propto
\exp (-c/\alpha_W)$, with $0~<~c~\lwig~2\pi$.
However, this does not exclude the possibility
of a small coefficient $(c\ll 1)$ such that an observable
cross section results$^{18}$.
The lower limit of
$\hat \sigma_0 $ is characteristic of
a geometrical ``weak'' cross section and the upper limit
range is a geometrical ``strong''
cross section suggested by analogies between the weak
SU(2) gauge sector and
the color SU(3) gauge sector$^{19}.$

The simultaneous production of ${\cal O}(30)$ W~bosons
at future had\-ron colliders such as the proposed
LHC would lead to
spectacular signatures$^{20,21}$.
Since  approximately 20  charged hadrons (mainly $\pi^\pm$'s) arise from
hadronic W decays one
could typically expect
400 $\pi^\pm$'s
in one multi--W event accompanied by $\simeq 400$ photons from the
decay of $\simeq 200$ $\pi^0$'s. The charged hadrons would
have a minimum average transverse momentum of order
$p_T^\pi \ge {\cal O}
(m_W/30 )\simeq (2-3)$ GeV
if the W bosons are produced without transverse momentum.
Similarly, one could expect $\simeq 5$
prompt muons ($\simeq 3$ from W decays
and $\simeq 2$ from $c$, $b$, or $\tau$ decay)
carrying a minimum average transverse momentum of
$p_T^\mu \ge {\cal O}( m_W/2 )\simeq 40$ GeV.
Analogous situations hold for other prompt leptons such as $e^\pm,
\nu$ etc.
No conventional reactions in the Standard Model are backgrounds to
multi--W processes$^{21}.$

Figure 1 shows the regions in $\sqrt{\hat{s}_0} - \hat \sigma _0$
space accessible to the LHC.
The contours give the number of multi-W reactions occurring
during $10^7$~s of operation (assuming 100\% detection efficiency).
These contours may be used as a benchmark to evaluate the
effectiveness of various cosmic ray physics experiments for constraining
multi--W phenomena$^{12}$.

\begin{center}
\epsfxsize=12cm
\hspace*{0in}
\epsfbox[54 119 538 543]{ 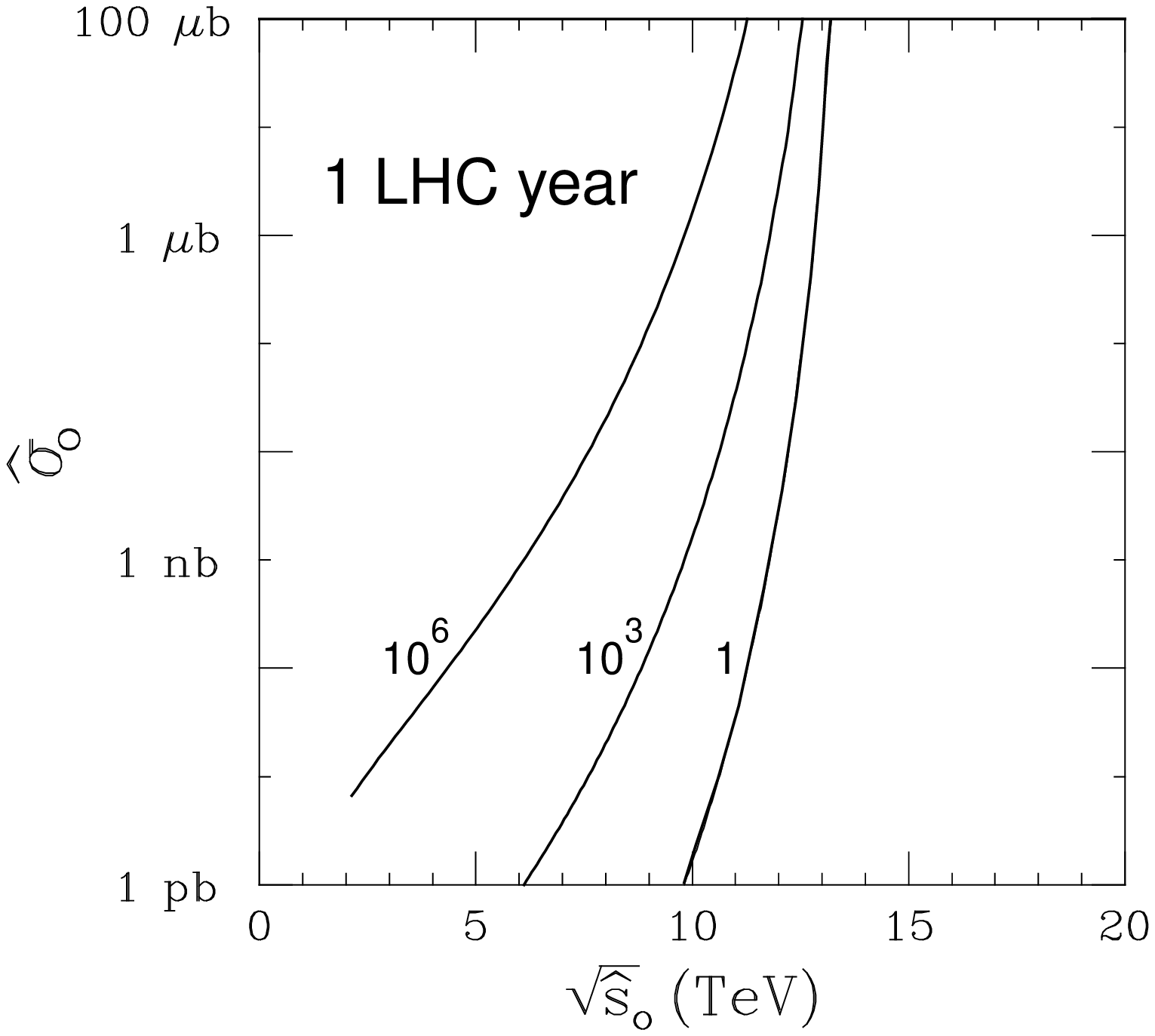       }

\parbox{12cm}{ \capfont
Fig. 1.  Contours corresponding to $1,10^3,10^6$ multi--W
events in one year (10$^7$~s) of operation for the LHC
($\sqrt{s_{pp}} = 14$ TeV; luminosity $10^{34}$ cm$^{-2}$s$^{-1}$).
 }
\end{center}

\vglue 0.6cm
{\elevenit\noindent 2.3. NESTOR's Multi--W(Z) Discovery Potential}
\vglue 0.4cm

Until the proposed LHC becomes available, cosmic rays provide our only
access to parton--parton center of mass energies in the multi--TeV
range. However, having energetic cosmic rays is not sufficient;
due to small event rates, it is only practical to search for phenomena
which have spectacular low--background signatures. In
this section we review the appeal of neutrino--induced phenomena
and discuss NESTOR's potential for investigating multi--W processes.

While there is an ample flux of multi--PeV
cosmic protons bombarding the Earth, it appears hopeless that
one could exploit this flux to search for multi--W phenomena$^{12}$.
The source of the difficulty is that, due to a generic QCD--dominated
${\cal O}(100~{\rm mb})$ proton--proton total cross section, only a small
fraction ($\sigma_{\rm multi-W}^{pp} / \sigma_{\rm total}^{pp}$) of the
incident cosmic proton flux is available for multi--W processes.
The proton flux attenuation due to large competing generic QCD cross sections
is so severe that even for the most optimistic values of $\sqrt{\hat{s}_0}$
and $\hat{\sigma}_0$ only 1--100 extensive air showers of
multi--W origin would be incident on $100~{\rm km}^2$ in one year
(see Ref.~12 for details). Moreover, the characteristics of these
1--100 multi--W showers would not be sufficiently distinctive
to avoid confusion with fluctuations in the background of
generic air showers. Given these prospects, one abandons hope
of exploiting the cosmic proton flux for uncovering multi--W processes.

Cosmic neutrinos, on the other hand, suggest an attractive
alternative. Since conventional charged current neutrino--nucleon
cross sections are anticipated to be of ${\cal O}(1-10$~nb)
in the energy range in which we are interested, neutrino flux
attenuation due to generic processes is not an issue
unless we wish to search for phenomena with cross sections
orders of magnitude smaller. Furthermore, we gain the
added discrimination that neutrino--initiated phenomena
typically occur deep in the atmosphere or inside the Earth
due to a small total neutrino--nucleon cross section (which
conceivably could be dominated by new physics).

There is, however, a drawback to using ultrahigh energy
neutrinos as a probe for new physics: we do not know their flux.
Though atmospheric neutrinos ({\it i.e.,} neutrinos produced in hadronic
showers in the atmosphere) are a guaranteed source
of neutrinos, their flux in the PeV region is anticipated
to be negligible$^{22}$ (see Fig. 2).
More promising are recent predictions
of a sizeable flux of PeV neutrinos from active galactic
nuclei (AGN)$^{23-26}$. For definiteness, we
consider the (revised) Stecker {\it et al.}$^{23}$ AGN neutrino flux
of Fig.~2. AGN neutrino fluxes calculated under different assumptions
in Refs.~24,~25 generally agree with Ref.~23
above .1~PeV, which is the energy range we are interested in.
In this sense our use of the Stecker {\it et al.} flux
is intended to be representative of a large class
of AGN flux models. In Ref. 12
we have checked that,
within the parameter ranges of Eq.~(\ref{eq:partoncro}),
large neutrino
cross sections for multi--W production are consistent with
proposed AGN flux models.

In addition to the proposed AGN neutrino fluxes, we will also consider
the possibility of exploiting neutrinos produced when
ultrahigh energy protons inelastically scatter off the cosmic
microwave background
radiation$^{27}$ (CBR) in processes
such as $\gamma p \rightarrow \Delta
\rightarrow n \pi^+$ where the produced pion
subsequently decays$^{28-30}.$ As shown in Fig.~2,
such processes may provide the
dominant component of the neutrino flux at
energies beyond $\simeq 1$ EeV.
The CBR neutrino fluxes shown in Fig.~2
are taken from Ref. 23.

\begin{center}
\epsfxsize=13cm
\hspace*{0in}
\epsfbox[42 42  552 495      ]{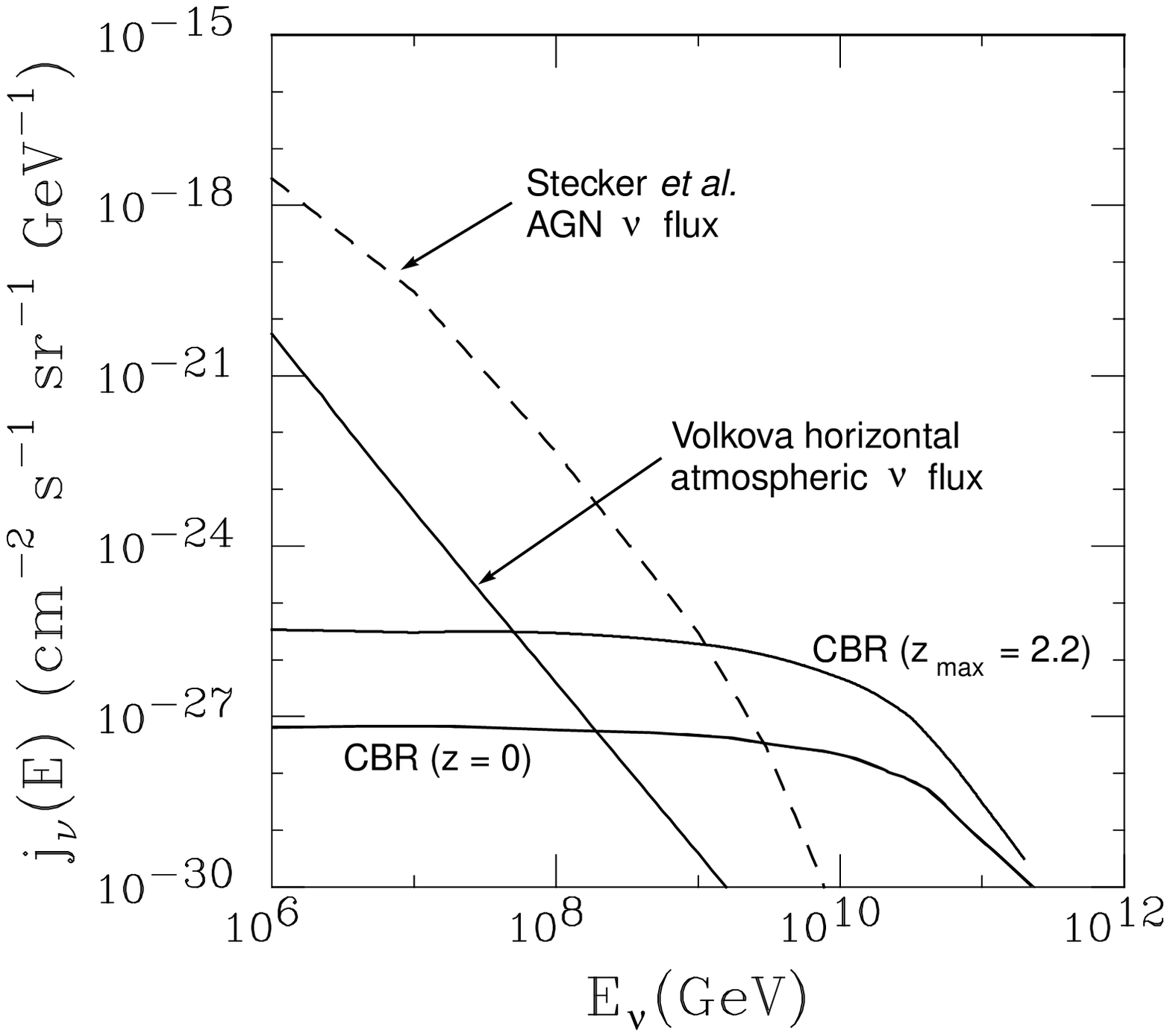      }

\parbox{12cm}{ \capfont
Fig. 2. Differential flux of neutrinos used in text.
The horizontal flux of atmospheric neutrinos is taken from
Volkova$^{22}$. The diffuse flux of neutrinos from
active galactic nuclei (AGN) and the neutrino flux due
to inelastic scattering of cosmic ray protons off the cosmic
microwave background radiation (CBR) are from
Stecker {\mbox{\tenit et al.}}$^{23}$. The above fluxes are summed
over all neutrino species.
 }
\end{center}

NESTOR may be used to search for multi--W phenomena in at least two
ways: 1) through the detection of muon bundles which
originate from the prompt decays of many $W$ bosons produced far
$ (\gwig {\cal O}( 1~{\rm  km} )) $ from the
detector$^{12-15}$ and 2) through the detection of
the cascade produced by a nearby  $( \lwig {\cal O}(1~{\rm km }))$
neutrino--nucleon multi--W process. For
definiteness, we will restrict our considerations to
the stage--2 NESTOR  array depicted in
Fig. 3. A stage--1 NESTOR array consisting of a single tower (as
in Fig.~3c) is comparable to a fully constructed DUMAND array
whose response to multi--W phenomena is discussed in Ref.~12.

\begin{center}
\epsfxsize=15cm
\hspace*{0in}
\epsfbox[14 42 581 675]{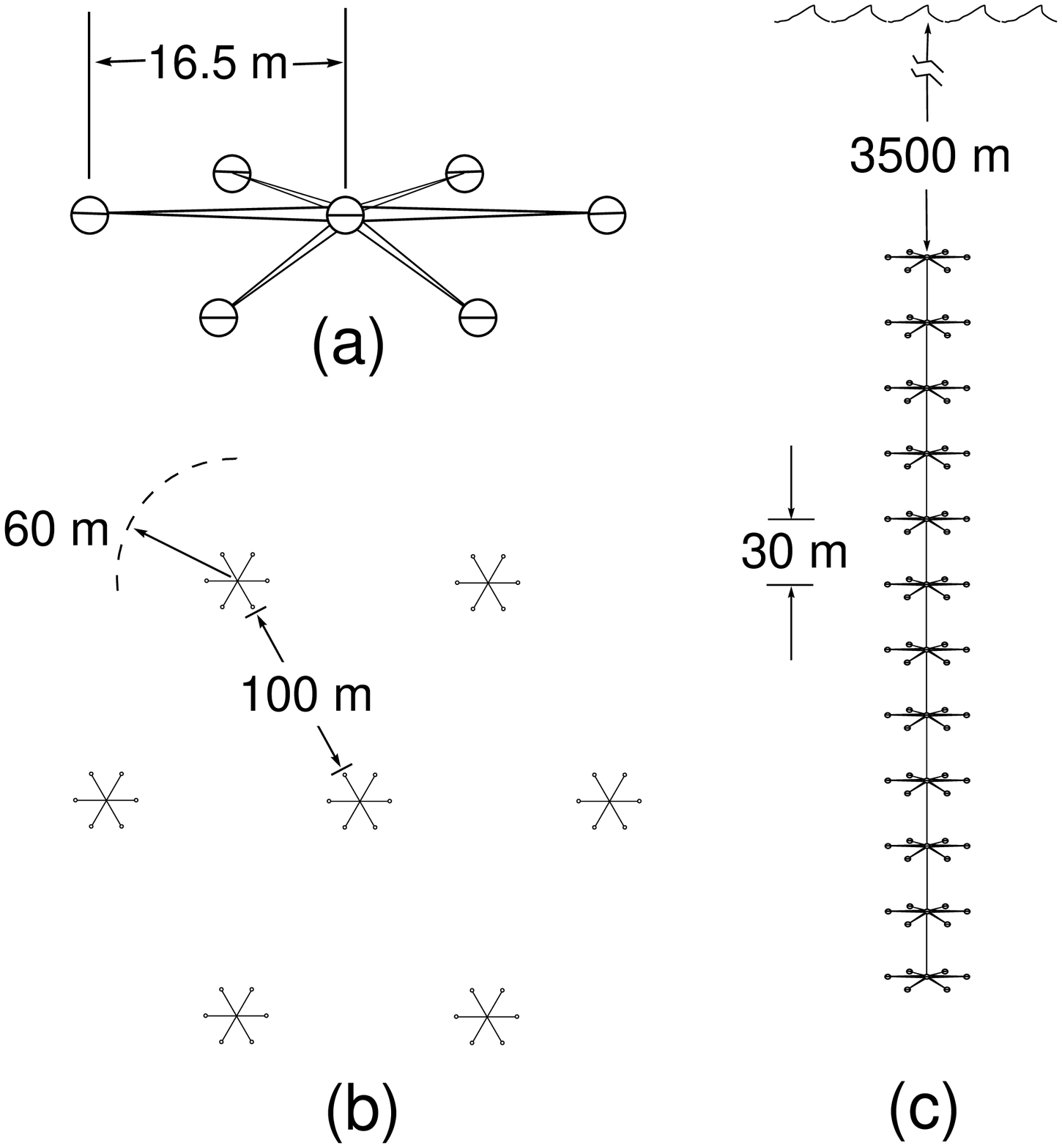}

\parbox{13cm}{ \capfont
Fig. 3. Parameters of the stage--2 NESTOR  array assumed in this
paper. (a)~Hexagonal frame of 16.5~m radius supports
up-- and down--looking phototubes. (b)~Top view of stage--2 array
composed of seven towers. For simulations
we assume each tower efficiently reconstructs muons
within 60 m radius (appropriate for 10~TeV muons).
(c)~A single tower is made up of twelve hexagonal frames.
The average sea depth of a tower is 3665~m ( = 3500~m + 330/2~m ). }
\end{center}

Figure 4a shows the expected number of multi--W events detected
through muon bundles in $10^7~{\rm s}$ for a stage--2 NESTOR
array assuming the AGN neutrino flux of Stecker {\it et al}.
Details of the calculation may be found in Refs.~12,13.
As a simplifying approximation, we assume that the stage--2
NESTOR array reconstructs muon trajectories inside a
cylindrical fiducial volume of radius 193~m ( = 60~m + 100~m +
2 $\times$~16.5~m from Figs.~3a,b) and
height of 330~m (from Fig. 3c). The muon bundles consist
of 2--3 muons originating from the prompt decays of 30 W bosons.
Also shown in Fig. 4a are contours of
the average zenith angle cosine corresponding to the
arrival direction of the muon bundles.

\vspace{.3cm}
\begin{center}
\epsfxsize=13cm
\hspace*{0in}
\epsfbox[22 76  603 600 ]{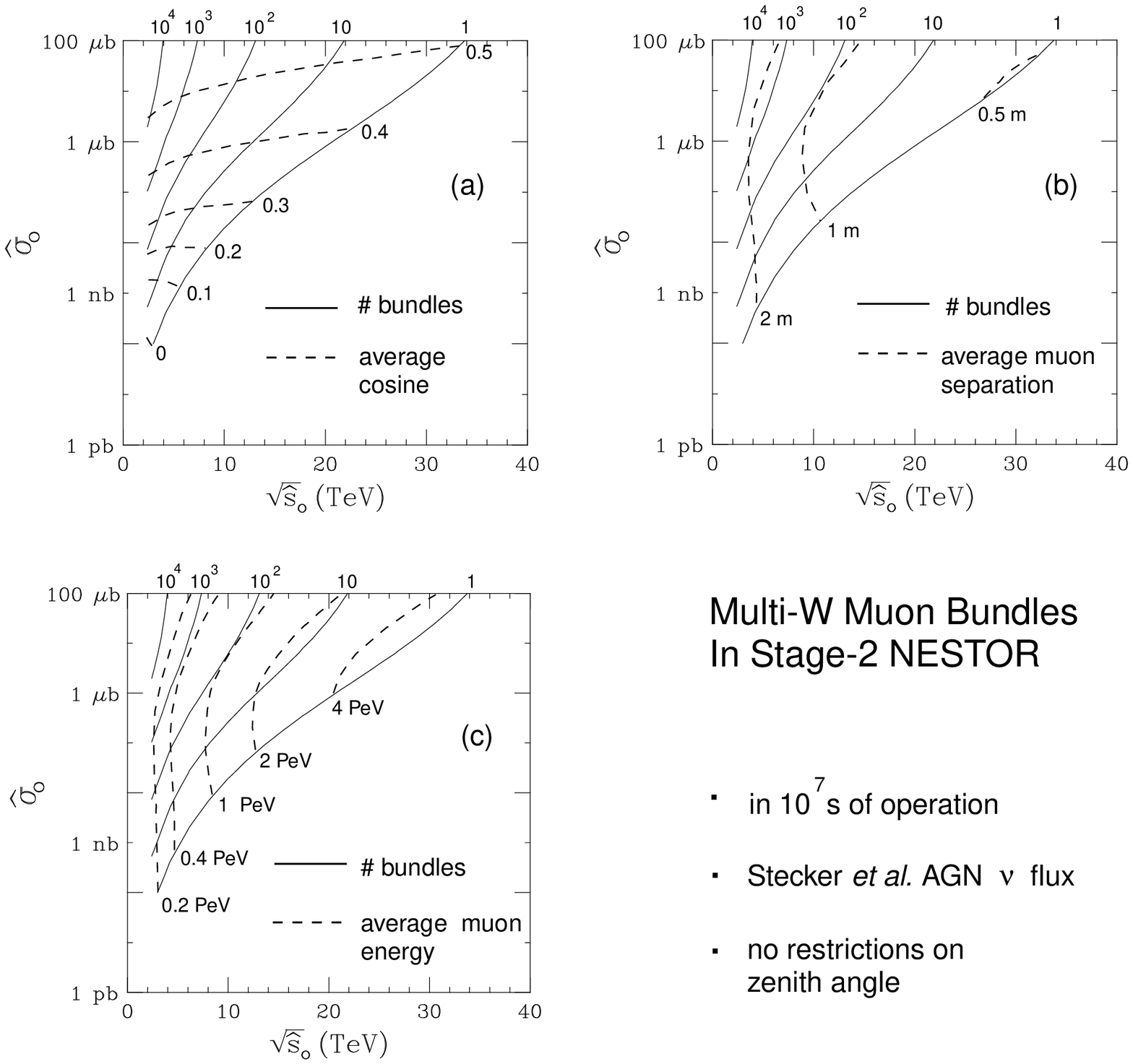}

\parbox{12cm}{ \capfont
Fig. 4. a) Contours for the number of muon bundles from
neutrino--initiated multi--W processes passing through
a stage--2 NESTOR fiducial volume in $10^7$~s assuming
the Stecker {\mbox{\tenit et al.}} AGN neutrino flux. Solid contours
give the number of events (labels appear along top of graph)
while dashed contours indicate the average zenith angle cosine of the
muon bundle arrival direction.
b) Solid: same as in (a);
dashed: average inter-muon separation within a muon bundle.
c) Solid: same as in (a);
dashed: average muon energy as muons pass through fiducial volume.}
\end{center}

        Figure~4b shows the average inter--muon separation
within a muon bundle of multi--W origin. Inter--muon separations
of less than a few meters call into question the ability of
a water Cherenkov detector to recognize the multi--muon
nature of such events. To our knowledge this experimental
issue has not been addressed in any detail. Figure~4c
includes contours of the average muon energy as the muons
pass through the detector. Muon energies in the
range of 10--1000~TeV ensure their detection. The large muon
energies (and small inter--muon separations) of Figs.~4b,c
reflect the kinematics of the enormous cosmic
neutrino energy required to initiate a multi--W process
characterized by a multi--TeV parton--parton threshold.

        We emphasize that the contours in Fig.~4 only
chronicle the {\it average} value of each observable at each
point in $\sqrt{\hat{s}_0}-\hat{\sigma}_0$ space. For example,
Fig.~5 shows the distributions of zenith cosine, intermuon
separation and muon energy for
the parameter choice ($\sqrt{\hat{s}_0}=4~{\rm TeV}$,
$\hat{\sigma}_0=10~{\rm nb}$) --- the averages of these
distributions correspond to single points in Figs.~4a,b,c respectively.

\begin{center}
\epsfxsize=15cm
\hspace*{0in}
\epsfbox[29 294 597 597  ]{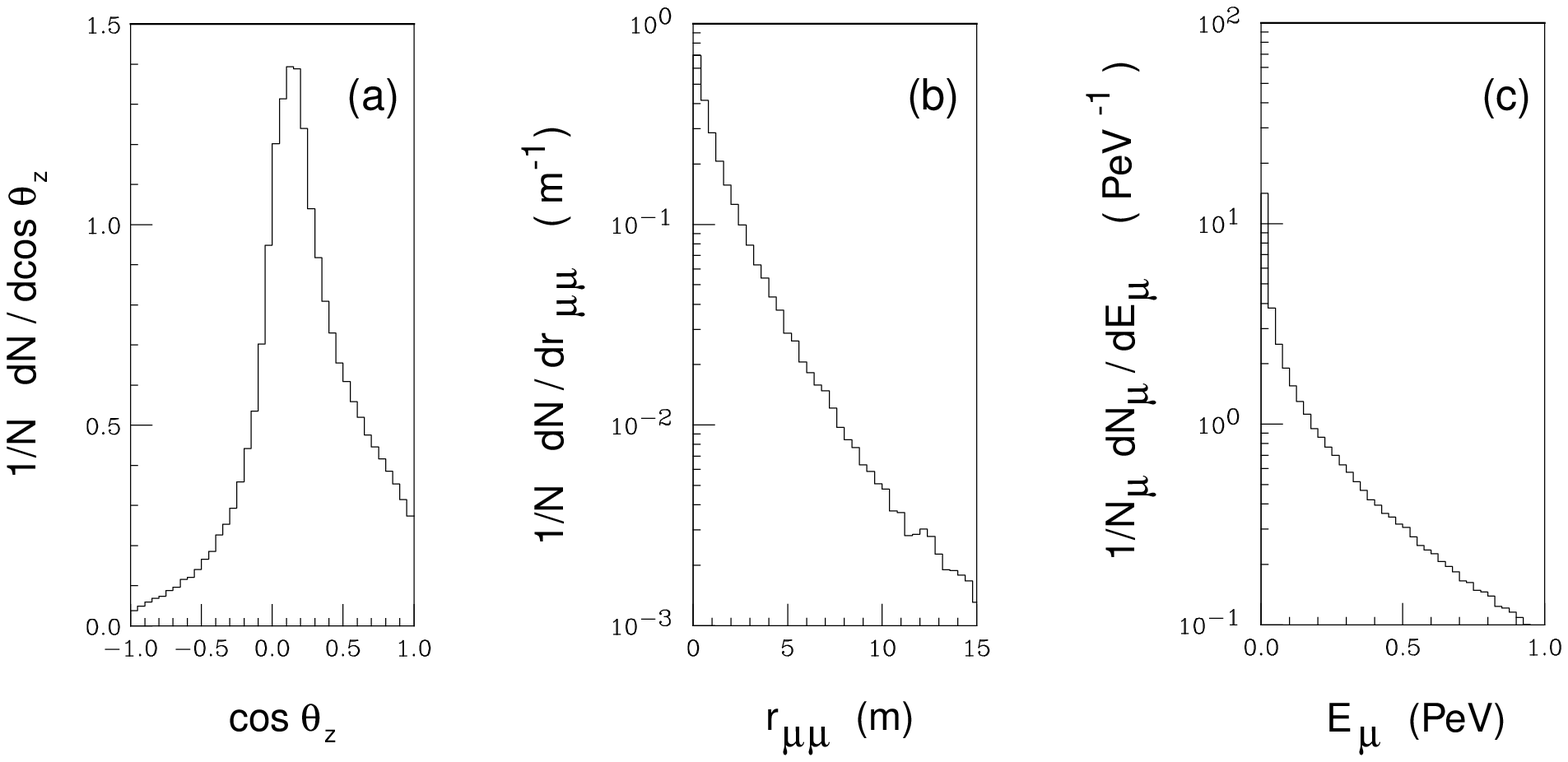}

\parbox{15cm}{ \capfont
Fig. 5. Distributions corresponding to muon bundles from
neutrino--initiated multi--W processes (with fixed
$\sqrt{\hat{s}_0} = 4~{\rm TeV}$, $\hat{\sigma}_0 = 10~{\rm nb}$)
passing through a NESTOR stage--2 fiducial volume  assuming
the Stecker {\mbox{\tenit et al.}} AGN neutrino flux. (a) Distribution of
muon bundle arrival direction ($\langle \cos \theta_z \rangle = .21 $).
(b) Distribution of the average lateral separation between muons
within a bundle. Assuming the production of exactly $3$ muons from
the prompt decays of $30$ W bosons, approximately $90\%(10\%)$ of
the bundles making it to the detector consist of $3(2)$ muons
( $\langle r_{\mu \mu} \rangle = 2.0~{\rm m}$).
(c) Distribution of individual muon energy as muon passes through
fiducial volume ($\langle E_{\mu} \rangle = .33~{\rm PeV}$).
Distributions shown in (a),(b) and (c) are normalized to unity.
A total of 22 muon bundles pass through the fiducial volume
in $10^7$~s.
}
\end{center}

To eliminate the possibility of a muon--bundle background
due to atmospheric muons produced in generic hadronic air showers,
one can restrict a search to muon bundles arriving only from zenith angles
with $\theta_z > 80^o.$ By looking towards large zenith angles we are assured
that muons reaching the detector are produced inside the Earth and hence
are not due to generic interactions in the atmosphere. Figure~6
illustrates the potential of a stage--2 NESTOR which looks only at
muon bundles with $\theta_z > 80^o.$

\begin{center}
\epsfxsize=13cm
\hspace*{0in}
\epsfbox[17 80  599 662]{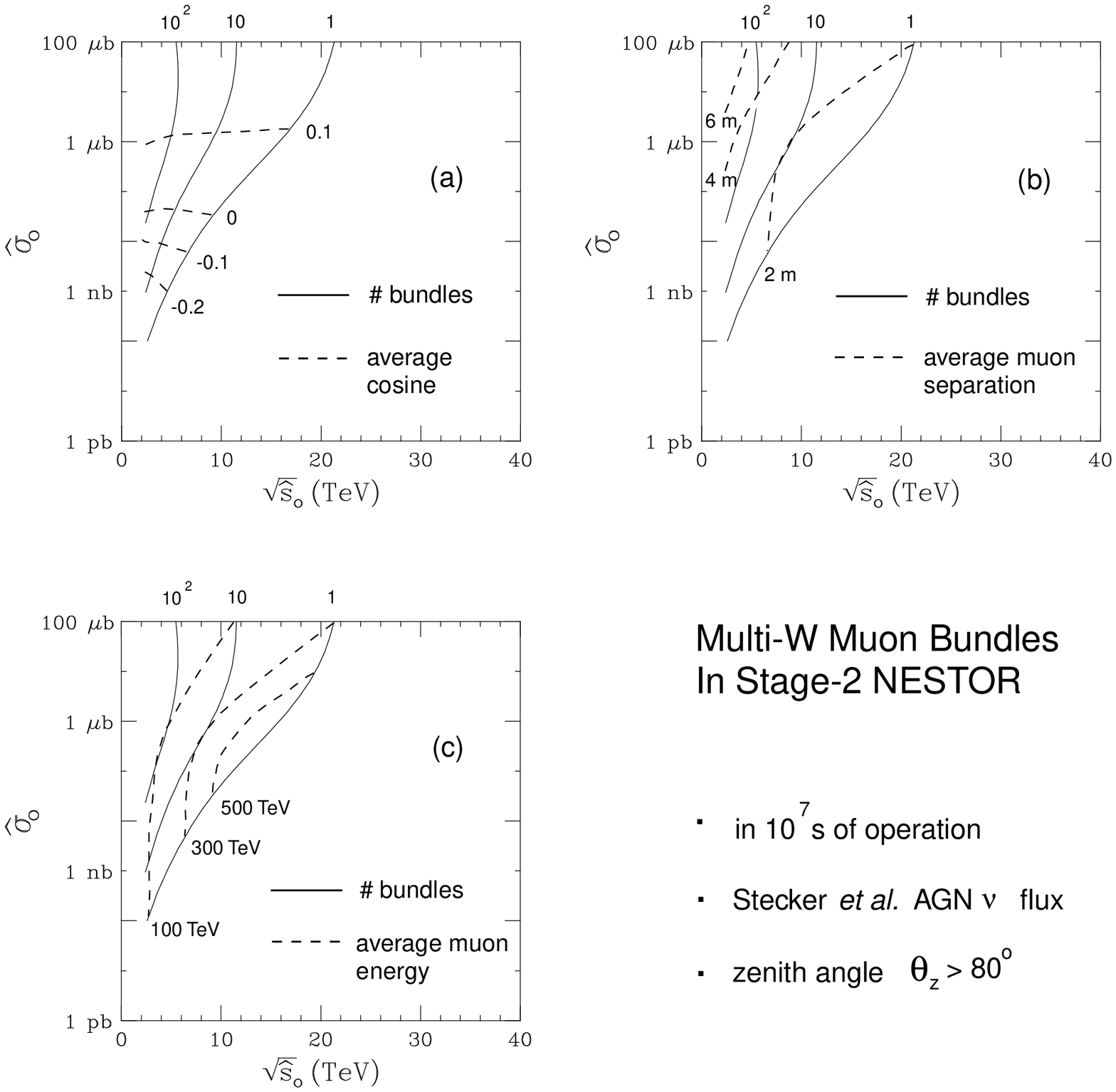}

\parbox{12cm}{ \capfont
Fig. 6. a) Contours for the number of muon bundles arriving
with zenith angle $\theta_z > 80^o$ from
neutrino--initiated multi--W processes passing through
a stage-2 NESTOR fiducial volume in $10^7$~s assuming
the Stecker {\mbox{\tenit et al.}} AGN neutrino flux.
Solid contours give the number of events while dashed contours
indicate the average zenith angle cosine of the
muon bundle arrival direction.
b) Solid: same as in (a);
dashed: average inter-muon separation within a muon bundle.
c) Solid: same as in (a);
dashed: average muon energy as muons pass through fiducial volume.}
\end{center}

If, in addition to AGN neutrinos, we include a
flux component due to neutrino production off the cosmic microwave
background, we obtain the discovery contours  for near--horizontal
muon bundles shown in Fig.~7. While it appears that
including a CBR neutrino flux component can,
in principle, greatly extend the range
of $\sqrt{\hat{s}_0}$ probed, we must point out the sensitivity to
the flux model assumed. For example, had we used the minimum
CBR neutrino flux (labelled $z=0$ in Fig.~2) the contour corresponding
to one detected event in Fig. 7, which extends up to
$\sqrt{\hat{s}_0} \simeq 50~{\rm TeV},$ would collapse back to
its position shown in Fig.~6.

\begin{center}
\epsfxsize=13cm
\hspace*{0in}
\epsfbox[14 110 605 665]{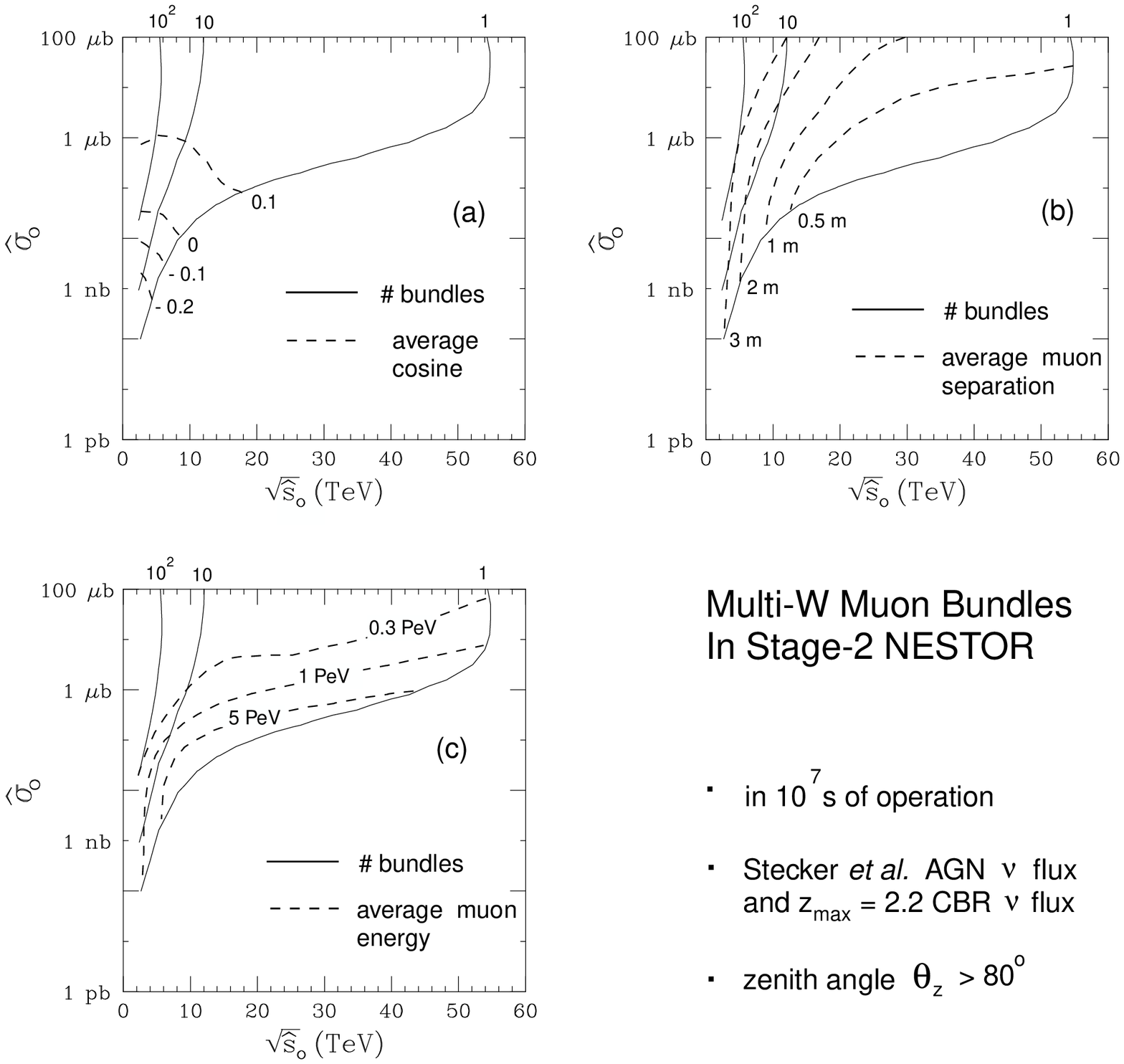}

\parbox{12cm}{ \capfont
Fig. 7. a) Contours for the number of muon bundles arriving
with zenith angle $\theta_z > 80^o$ from
neutrino--initiated multi--W processes passing through
a stage--2 NESTOR  fiducial volume in $10^7$~s assuming
the Stecker {\mbox{\tenit et al.}} AGN neutrino flux and the CBR
neutrino flux component labelled $z_{\rm max}=2.2$ in Fig.~2.
Solid contours give the number of events while dashed contours
indicate the average zenith angle cosine of the
muon bundle arrival direction.
b) Solid: same as in (a);
dashed: average inter--muon separation within a muon bundle.
c) Solid: same as in (a);
dashed: average muon energy as muons pass through fiducial volume.}
\end{center}

Turning now to the detection of underwater cascades, the
effective volume of stage--2 NESTOR can be much larger than
the volume appropriate for detecting the Cherenkov light from
throughgoing muons; we will assume a sensitive volume
of 1~km$^3$, appropriate for multi--PeV cascades,
for ``contained'' multi--W phenomena. Figure  8
shows the discovery contours for contained multi--W phenomena
assuming the Stecker {\it et al.} AGN neutrino flux and
a CBR neutrino flux component (labelled $z_{\rm max}=2.2$ in Fig.~2).
The contour for one detected event in Fig.~8 is subject
to the same sensitivity to the CBR flux model as
discussed in the previous paragraph. The contours for the average
energy of a multi-W cascade assume that $20\%$ of the initiating
neutrino's energy is given to prompt neutrinos and muons (through
W decay) and is not observed.

\begin{center}
\epsfxsize=15cm
\hspace*{0in}
\epsfbox[14 30 603 320 ]{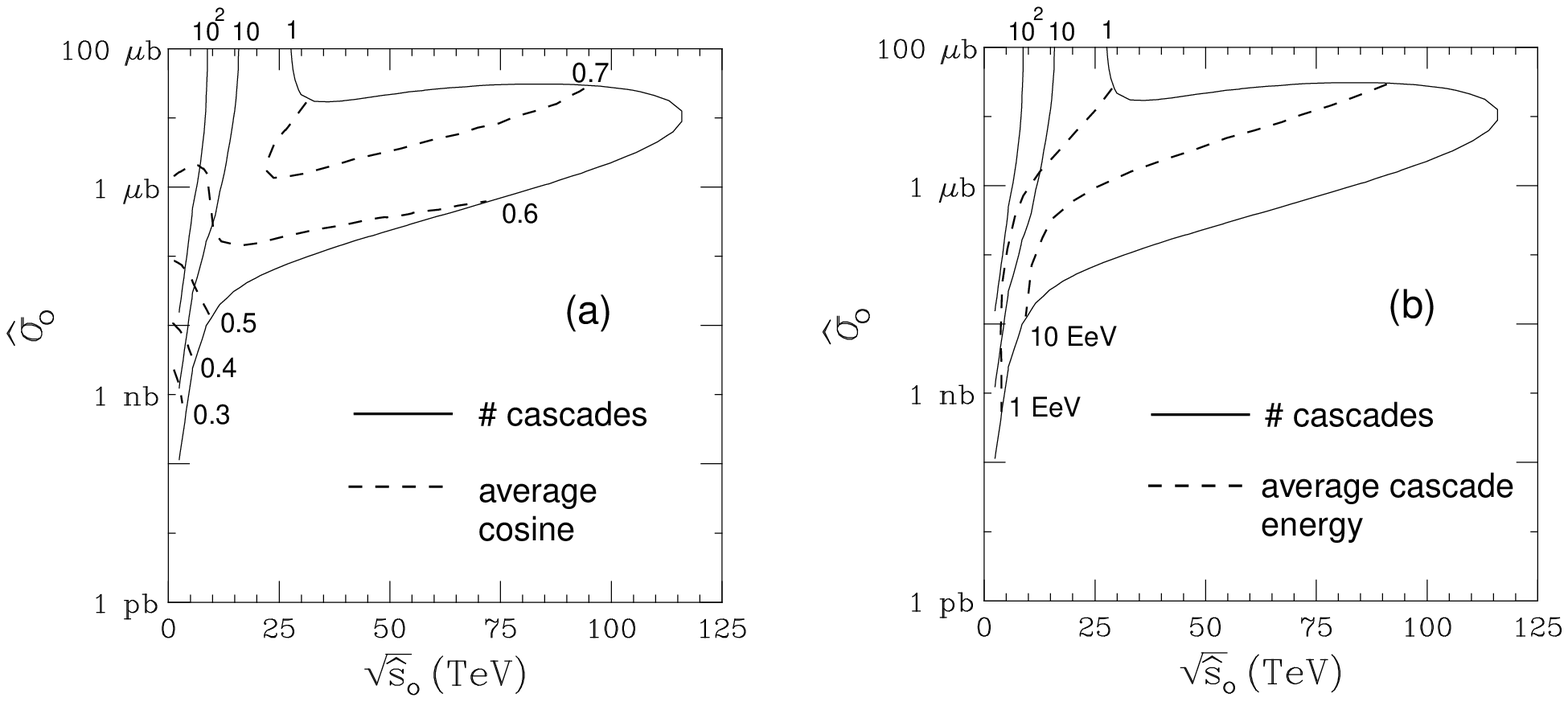}

\parbox{14cm}{ \capfont
Fig. 8. a) Contours for the number neutrino--initiated
multi--W processes contained in a 1~km$^3$ fiducial volume
at stage--2 NESTOR in $10^7$~s assuming
the Stecker {\mbox{\tenit et al.}} AGN neutrino flux and the CBR
neutrino flux component labelled $z_{\rm max}=2.2$ in Fig.~2.
Solid contours give the number of events while dashed contours
indicate the average zenith angle cosine of the original neutrino.
b) Solid: same as in (a); dashed:
average energy of the contained cascade.}
\end{center}

Unlike the search for near--horizontal muon bundles, contained
multi--W phenomena may have backgrounds due to generic
charged current processes. We estimate these backgrounds as
follows. At the energies we are interested in,
the lepton $\ell$ in a generic charged current process
$\nu N \rightarrow \ell + X$    carries away an average
of 80\% of the incident neutrino energy$^{31}$. Hence if
$\ell$ is a muon, only about 20\% of the incident neutrino energy
can appear in the contained cascade. Treating generic
$\nu_{\mu}$ and $\bar{\nu}_\mu$ cascades in this manner, we
consider the background to contained multi--W production
to be any contained cascade with energy greater than $E_{\rm thresh}/10$
where $E_{\rm thresh}$ is the  threshold energy required
of a neutrino to initiate a multi--W process. We include background
cascades with energies ten times less than $E_{\rm thresh}$
in an attempt to account for the limited energy resolution of
neutrino telescopes.
Figure~9  compares the rate, energy and
direction of contained cascades due to generic charged current
interactions and  multi--W processes.

\begin{center}
\epsfxsize=14cm
\hspace*{0in}
\epsfbox[17 150 603 719]{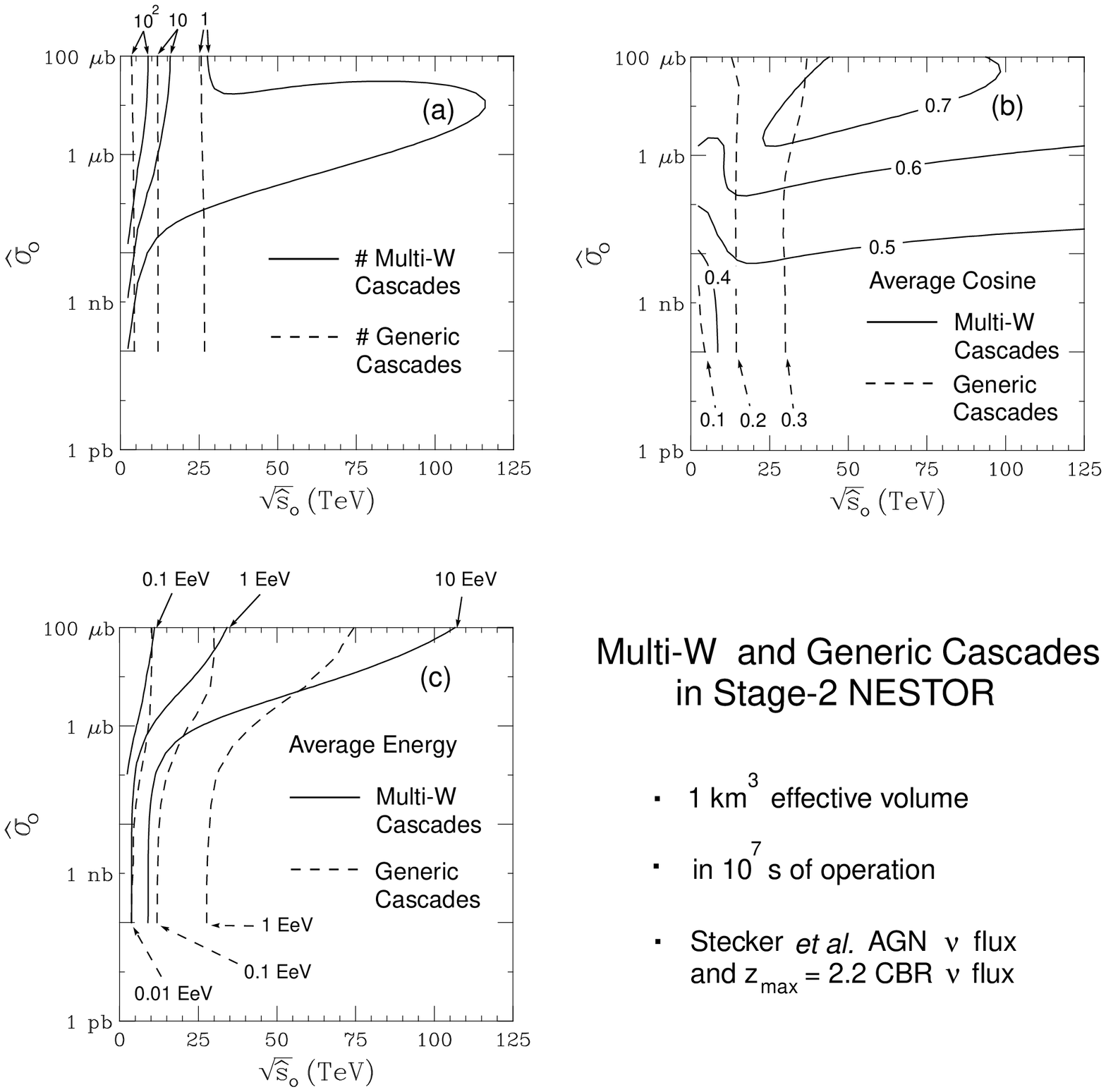}

\parbox{14cm}{ \capfont
Fig. 9. Comparison of  a) rate b) direction and c) energy
of contained cascades due to multi--W processes  (solid)
and generic charged current interactions (dashed).
Contours correspond to neutrino--initiated cascades in
a 1~km$^3$ fiducial volume
at stage--2 NESTOR in $10^7$~s assuming
the Stecker {\mbox{\tenit et al.}} AGN neutrino flux and the CBR
neutrino flux component labelled $z_{\rm max}=2.2$ in Fig.~2.
Multi--W contours are same as in Fig.~8. Criteria for
background cascades are described in text.}
\end{center}

An additional process which
will lead to spectacular contained cascades is due to resonant
W production through the Glashow process
$\overline{\nu}_e + e^{-} \to W^{-} \to$
hadrons. Under the same detector and flux conditions as in Fig.~9,
we find that there should be
approximately 9 contained resonant interactions
in $10^7$~s at stage--2 NESTOR of which 6 would involve hadronic
decays of the $W$ boson. These 6 cascades would have
average energies
of 6.3~PeV and hence would not be a serious background to
multi--W phenomena which are characterized by much higher cascade
energies.

For completeness, we have checked the multi--W discovery
potential of stage--2 NESTOR subject only to the existence
of atmospheric neutrinos or
the minimum flux of neutrinos due to protons scattering
inelastically off the cosmic microwave background (labelled $z=0$
in Fig.~2). Only if $\sqrt{\hat{s}_0} \lwig 3~{\rm TeV}$ and
$\hat{\sigma}_0 \gwig 10~\mu{\rm b}$ could stage--2 NESTOR  expect
to see more than one contained multi--W cascade in $10^7$~s due to
atmospheric neutrinos. Even less promising is the minimum flux
due to the CBR which is at least two orders of magnitude too small
to be useful to stage--2 NESTOR in any region of multi--W parameter
space.

\begin{center}
\epsfxsize=10cm
\hspace*{0in}
\epsfbox[51 147 519 595 ]{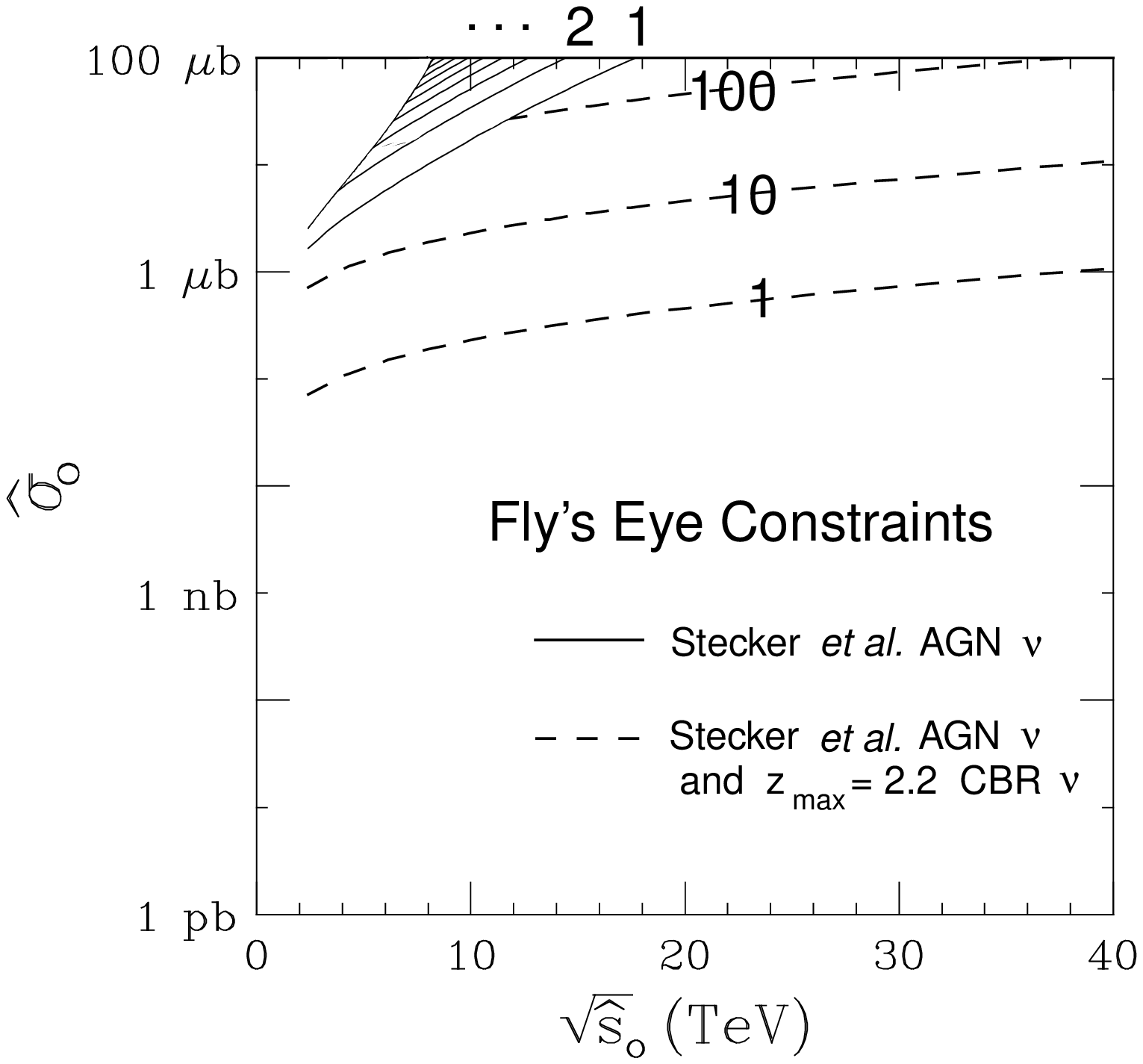}

\parbox{12cm}{ \capfont
Fig. 10. Number of multi--W events expected by Fly's Eye assuming
a) only the Stecker {\mbox{\tenit et al.}} AGN neutrino flux (solid contours
labelled for 1,2, ... events) and b) sum of
the Stecker {\mbox{\tenit et al.}} flux and
neutrinos produced off the cosmic background radiation (dashed
contours). }
\end{center}

As a final point regarding multi--W processes, we
present in Fig.~10 the current best constraints
on multi--W phenomena which come from the Fly's Eye experiment
(see Ref. 12 for a discussion). Figure~10 supersedes
the Fly's Eye constraints presented in Refs.~11,~12;
a recent re--evaluation$^{32}$ of the Fly's Eye data uncovered
an analysis error which overestimated
the Fly's Eye's sensitivity to air showers
initiated deep in the atmosphere. Consequently, many of the results
presented in Ref.~33 (on which we based our earlier constraints)
are optimistic by roughly an order of magnitude.
 Fly's Eye constraints on multi--W phenomena which assume
only atmospheric neutrinos or the minimum CBR neutrino flux ($z=0$
in Fig.~2) are orders of magnitude weaker than those presented in Fig.~10
and hence we do not present them.

\vspace{.5cm}
{\elevenbf\noindent 3. NESTOR's Sensitivity to
Compositeness}
\vglue 0.4cm
{\elevenit\noindent 3.1. Compositeness Scenarios}
\vglue 0.4cm
The main theoretical reasons for contemplating new physics beyond
the Standard Model are well known.
The striking generation pattern, {\it i.e.} the proliferation of
quarks and leptons and their deep inter-relations
remain
mysterious. Composite models of quarks and leptons may offer a
scenario for explaining these features$^{34}$.
There are a number of composite
models based on the hypothesis that (some) leptons and quarks
(and possibly even W's and Z's)
are bound
states of preonic constituents
characterized by a compositeness scale (inverse size) $\Lambda_c$.
However, there is not yet a
satisfactory model
in which the light masses of quarks and leptons can be
reconciled with their small radii $< {\cal O}(10^{-16}$ cm)
corresponding to a compositeness scale of 1 TeV and beyond.
Therefore we will only  exploit those
features of composite models which are fairly model independent.
Reference~35  has explored the features of neutrino-induced
air showers in the context of composite models
where hypothesized colored subconstituents of PeV neutrinos
interact with typical QCD cross sections$^{36}$.
\vglue 0.6cm
{\elevenit\noindent 3.2. Compositeness Parametrization}
\vglue 0.4cm
Typical features of composite models
are new interactions between quarks and leptons
and the occurrence of
novel particle species such as excited states towering  the
known lepton and quark ground states like leptoquarks and
leptogluons.

At low energies, $\sqrt{\hat s}\ll \Lambda_c$,
the new interactions are suppressed by inverse powers of the
compositeness scale.
The dominant effect should come from the lowest dimensional interactions
with four fermions (contact terms) whose most general chirally invariant
form reads$^{37}$
\[
 {\pm g \over 2 \Lambda^2 }
\Biggl[
\eta_{LL}\overline{\psi}_L\gamma_\mu
\psi_L\overline{\psi}_L\gamma^\mu\psi_L
+
\eta_{RR}\overline{\psi}_R\gamma_\mu
\psi_R\overline{\psi}_R\gamma^\mu\psi_R
+
2\eta_{LR}\overline{\psi}_L\gamma_\mu
\psi_L\overline{\psi}_R\gamma^\mu\psi_R
\Biggr] .
\]
\begin{equation}
\label{eq:contact} \qquad
\end{equation}
Such interactions can arise by constituent interchange and/or
by exchange of the binding quanta.
Various accelerator limits on the scale $\Lambda$
may be placed
through the non--observation$^{38}$ of virtual
interference effects of
 the new couplings, Eq. (\ref{eq:contact}),
with Standard Model contributions, {\it e.g.}
$\Lambda^{+(-)}_{LL}(eeqq) > 1.7(2.2)$ TeV,
$\Lambda^{+(-)}_{LL} (ee\mu\mu ) > 4.4(2.1)$ TeV
(here we used the conventions of the Review of Particle Properties$^{38}$).
The proposed LHC may enlarge such limits to
$\sim 20$ TeV for the compositeness scale probed in quark--quark
scattering$^{39}$.
 At present there are no
stringent bounds for neutrino--quark
couplings.
The {\it direct} contributions of the new couplings, Eq. (\ref{eq:contact}),
to scattering cross sections, being proportional to
${\hat s}/\Lambda^4$, are subdominant when compared to the {\it virtual}
contributions and therefore negligible at present collider energies.

At energies above the compositeness scale $(\sqrt{\hat s} \geq \Lambda_c)$,
the power suppression of the new interactions disappears and
a model independent
contact interaction analysis ceases to be useful. However,
on general grounds
one expects$^{40}$, in analogy
to strong interactions, that the neutrino--quark inelastic
cross section saturates, up to
logarithmic variations, at a geometrical value set by the
compositeness scale,
\begin{equation}
\label{eq:compsat}
{\hat\sigma}_c (\sqrt{\hat s} \geq \Lambda_c )
\simeq
{\pi\over \Lambda_c^2} ,
\end{equation}
where $\Lambda_c$, which characterizes the physical size of the
composite neutrinos and quarks, is
expected to be related by a factor of
${\cal O}(1)$ with the model independent parameters defined in
 Eq. (\ref{eq:contact}). The final state should  be dominated
by multi--quark and multi--lepton production.

At neutrino telescopes, the majority of the exotic
neutrino--quark scattering processes due to compositeness will
come from the center of mass
energy region above the compositeness scale.
Correspondingly,
we neglect virtual interference effects and direct
effects at energies below the compositeness scale
which are suppressed by inverse powers of $\Lambda_c.$
We are thus led to the following simple--minded  parametrization
of the inelastic cross section of neutrino--quark scattering
due to compositeness,
\begin{equation}
\label{eq:compparam}
{\hat \sigma}_c = {\pi\over \Lambda_c^2}\
\Theta ( \sqrt{\hat s} - \Lambda_c ) ,
\end{equation}
which should be sufficient for our purpose.
Furthermore we neglect  possible
${\hat s}$--channel resonances (leptoquarks). For simplicity, we
assume that all quarks and leptons couple with the same
strength.

\vglue 0.6cm
{\elevenit\noindent 3.3. NESTOR's Compositeness Discovery Potential}
\vglue 0.4cm
Since we have no prediction for prompt multi--muons from
composite interactions, we concentrate on detecting
contained hadronic and electromagnetic cascades.
In Table~1 we present the characteristics of cascades
due to compositeness  at NESTOR
(assuming $10^7$s in a 1 km$^3$ volume)
for the Stecker {\it et al.} AGN neutrino flux and
the $z_{\rm max}=2.2$ CBR neutrino flux. A quick comparison
between the signal and background event rates (with
the background rate calculated as in Sect.~2.3) indicates
the difficulty of uncovering composite phenomena
above $\Lambda_c \gwig 1~{\rm TeV}.$ With the same
neutrino flux not even the Fly's Eye can place
meaningful limits on compositeness due to the small cross
sections involved.

\begin{center}
\parbox{12cm}{ \capfont
Table 1.
Characteristics of cascades initiated by neutrino--quark inelastic
scattering due to compositeness. The number of cascades refer to
1 km$^3$ fiducial volume at stage--2 NESTOR in 10$^7$ s assuming
the Stecker {\mbox{\tenit et al.}}
AGN neutrino flux plus the CBR neutrino flux
labelled $z_{\rm max}=2.2$ in Fig. 2. Characteristics of background
cascades are given in parenthesis.
}
\end{center}

\begin{center}
\begin{tabular}
  { | c |  c | c | c |  } \hline
               &                 &                 &                   \\
$\Lambda_c$    &  $\#$ of Cascades & $\langle E_{\rm cascade}\rangle$ &
                 $\langle \cos\theta_z\rangle$ \\
               &                 &                 &                   \\
\hline
               &                &                 &                   \\
1 TeV   &  120 (360)            &   120 (3) PeV   &   .25 (.09)        \\
               &                &                 &   \\
3 TeV   &  .4  (140)            &    1900 (9) PeV   &  .25  (.10)           \\
               &                &                 &                    \\
5 TeV   &   .03 (65)            &          7100 (19) PeV
         &  .33 (.12)    \\
                                &                 &                &    \\
\hline
\end{tabular}
\end{center}

{\elevenbf \noindent 4. Leptoquark Searches at NESTOR? \hfil}
\vglue 0.4cm
Leptoquarks are color triplet bosons with spin 0 or 1, non--zero
lepton and baryon number and fractional electric charge, which couple
to lepton--quark pairs. Their appearance is predicted in almost any
theoretical scheme beyond the Standard Model which aims at
establishing a connection between quarks and leptons,
be it by arranging them within common multiplets or by assigning
to them common substructure$^{41}$.
Clearly, $ep$ colliders such as HERA
($\sqrt{s_{ep}}\simeq 300$ GeV)
or LEP--LHC ($\sqrt{s_{ep}}\simeq 1$ TeV)
represent ideal machines for
hunting leptoquarks$^{42}$,
since they give rise to dramatic formation peaks in the
$eq$ subprocess at a fixed value of $x=m_{LQ}^2/s$.

An exhaustive classification of leptoquarks, whose interactions are
B and L conserving, flavor conserving and invariant
with respect to the Standard
   Model transformations and flavor has been performed in
Ref.~43.
 A comprehensive study of indirect bounds on scalar leptoquarks
may be found in Ref.~44. The bounds arising from low energy data
turn out to be considerably stronger than the first results from the
direct searches at HERA$^{45}$.

Bergstr\"om, Liotta and Rubinstein$^{14}$ have discussed the
possible effects of scalar leptoquarks at neutrino telescopes.
As an example,
they considered the production of a
scalar leptoquark with weak isospin zero,
$S_0$, whose interactions with the first generation
quarks and leptons are given in general by$^{43}$
\begin{equation}
\label{eq:leptoquark}
\lambda_L\overline{q}^c_L i\tau_2\ell_L S_0^\dagger
+\lambda_R\overline{u}^c_R e_R S_0^\dagger  .
\end{equation}
According to Eq.~(\ref{eq:leptoquark}), the leptoquark
$S_0$ will appear as
an $\hat s$--channel resonance in the $\nu_e d$ subprocess.
Assuming $\lambda_R\equiv 0
$\footnote{\ninerm\baselineskip=11pt
 Owing to constraints from pseudoscalar meson decays,
leptoquarks with masses of order 100 GeV can have sizeable couplings
only to either left-- or right--handed leptons$^{46}$.},
$S_0$ decays into $e+u$  and $\nu_e+d$ in the ratio
$1:1$. As noted in Ref.~14, events of this type may be detected
inside large underwater detectors as contained events where the
produced electrons will generate  multi--TeV cascades
($\langle E_{\rm cascade}\rangle \sim m_{S_0}^2/2m_p$).

 Optimistically assuming  a leptoquark mass of 100
GeV\footnote{\ninerm\baselineskip=11pt
The direct limit from CDF is $m_{S_0}>82$ GeV, for
$\lambda_R=0$$^{47}$.},
a coupling of the order of the electromagnetic coupling,
$\lambda_L=0.3$,
and the ``old" Stecker {\it et al.} AGN neutrino flux
(first reference in Ref.~23),
Bergstr\"om {\it et al.} predicted
about 40 contained downward--moving electron cascades from
leptoquarks per year in $2\times 10^6$~t in
\baselineskip=19pt
DUMAND~II\footnote{\ninerm\baselineskip=11pt
The effective volume of DUMAND~II for 1 TeV cascades is about
$1\times 10^7$ tons$^{48}$.}. This compares to about
20 events from generic $\nu_eN$ charged current
interactions. Scaling the leptoquark predictions by
$\simeq 1/45$ to account
for the revisions to the Stecker {\it et al.} AGN flux calculation
(c.f. second reference in Ref.~23)
one would expect 10 contained electron cascades per year
in $2.3\times 10^7$~t at stage--2 NESTOR. However, it should be noted that
recent indirect bounds give $\lambda_L< 0.03$,
for $m_{S_0}=100$ GeV, $\lambda_R=0$$^{44}$, which reduces the
number of electron cascades by another factor of $\sim (.03/.3)^2=.01$.
We thus conclude that the next generation of neutrino telescopes
cannot compete with HERA in the search for leptoquarks$^{42}$
with
masses in the range 100--300 GeV.

\vglue 1.6cm
{\elevenbf \noindent 5. Conclusions \hfil}
\vglue 0.4cm
It is clear that
experiments at
future supercolliders such as the
proton--proton collider LHC
and the electron--proton collider
LEP--LHC
offer the best prospects
for observing or
constraining exotic phenomena in quark--quark (LHC)
 and electron--quark
(LEP--LHC) scattering in the TeV center of mass  range,
such as
multi--W production, multi--quark and multi--lepton
production due to compositeness,
and leptoquark production.

Before the era of LHC  and LEP--LHC,
neutrino telescopes
such as AMANDA, Baikal NT--200, DUMAND, and NESTOR
suggest alternative strategies
to search for exotic scattering phenomena.
Neutrino--induced multi--W processes give rise to energetic
muon bundles and spectacular multi--PeV cascades. The
latter is true also for neutrino--induced multi--hadron and multi--lepton
production due to compositeness. Leptoquark production in
neutrino--quark scattering generates multi--TeV cascades for
leptoquark masses in the hundreds of  GeV range.
However,
the
accessible range in parameter space ({\it i.e.},
the subprocess threshold energy and the size of the
subprocess cross section for the multi--W and compositeness scenario,
the masses and the
couplings of the leptoquarks)
depends on the unknown ultrahigh energy neutrino flux.

\baselineskip=18pt
We have found that,  within the anticipated parameter range,
the prospects to constrain or observe multi--W
processes with the help of neutrino telescopes are superior to
searches for compositeness and
leptoquarks\footnote{\ninerm\baselineskip=11pt
This, however, may be misleading since
for multi--W
production we parametrized our ignorance by
two unrelated parameters, namely the
subprocess cross section, ${\hat\sigma}_0$, and the
threshold,
$\sqrt{{\hat s}_0}$, whereas for the compositeness scenario
we had only one parameter, the scale $\Lambda_c$, which determines both
the subprocess cross section ($\pi/\Lambda_c^2$) and the
threshold ($\Lambda_c$). A reduction of parameters for
multi--W  production by assuming that the
subprocess cross section
above threshold should be given by the geometrical size
of the sphaleron  (${\hat\sigma}_0 =\pi/{\hat s}_0$) or by the
$S$--wave unitarity bound (${\hat\sigma}_0=16\pi/{\hat s}$) would
lead to the same poor sensitivity as for compositeness.}.
If a sizeable diffuse flux of ultrahigh neutrinos exists,
{\it e.g.} at the level of the Stecker {\it et al.}
AGN neutrino flux and/or the
CBR neutrino flux labelled $z_{\rm max}=2.2$ in Fig. 2,
neutrino telescopes of the next generation may not only
be able to detect it through ordinary charged current interactions,
thereby establishing valuable constraints on multi--W parameter
space via the Fly's Eye limits, but may even indicate
whether multi--W processes are real or an artifact of our imperfect
understanding of multi--TeV weak interactions.
A sensible search for compositeness and leptoquarks would require
even larger neutrino fluxes or detectors larger than stage--2 NESTOR.

\vglue 0.5cm
{\elevenbf \noindent 6. Acknowledgements \hfil}
\vglue 0.4cm
One of us (A.R.) would like to thank Leo Resvanis for providing
details about stage--2 NESTOR and Wilfried Buchm\"uller for
discussions about compositeness and leptoquarks.
\vglue 0.5cm
{\elevenbf\noindent 7. References \hfil}
\vglue 0.4cm
\baselineskip=18pt
\elevenrm


\begin{thebibliography}{99}
\bibitem{amanda} S. Barwick {\it et al.}  (AMANDA Collaboration),
in {\it Proc. 26th Int. Conf. on High Energy Physics},
Dallas, August 1992, ed. J.R.~Sanford, (AIP, New York, 1993) Vol. 2,
p. 1250.
\bibitem{baikal}
I. Sokalski and Ch. Spiering (eds.)
(The Baikal Collaboration), Baikal preprint BAIKAL 92-03 (1992);
in {\it Proc. 23rd Int. Cosmic Ray Conf.}, Calgary, July 1993,
ed. D.A. Leaky, (U. of Calgary, 1993),
Vol. 4, p. 573; R. Wischnewski, in {\it these proceedings}.
\bibitem{dumand}
P. Bosetti {\it et al.} (DUMAND Collaboration),
Hawaii DUMAND Center preprint HDC-2-88 (1989);
C.M. Alexander {\it et al.},
in {\it Proc. 23rd Int. Cosmic Ray Conf.}, Calgary, July 1993,
ed. D.A. Leaky, (U. of Calgary, 1993),
Vol. 4, p. 515; P. Grieder, in {\it these proceedings}.
\bibitem{nestor} L.K. Resvanis (NESTOR Collaboration),
in {\it Proc. Workshop on
High Energy Neutrino Astrophysics}, Honolulu, March 1992,
eds. V. Stenger, J. Learned, S. Pakvasa and X. Tata,
(World Scientific, Singapore, 1993), p. 325; in
{\it these proceedings}.
\bibitem{ri90}
 A. Ringwald, {\it Nucl. Phys.} {\bf B330 }(1990) 1.
\bibitem{es90}
 O. Espinosa, {\it Nucl. Phys. } {\bf B343 }(1990) 310.
\bibitem{co90} J. Cornwall, {\it Phys. Lett.} {\bf B243 }(1990) 271.
\bibitem{gp90} H. Goldberg, {\it Phys. Lett. } {\bf B246 }(1990) 445.
\bibitem{go92a} H. Goldberg, {\it Phys. Rev.} {\bf D45} (1992) 2945.
\bibitem{ma92} M. Mattis, {\it Phys. Rep. } {\bf 214 }(1992) 159;
P. Tinyakov, {\it Int. J. Mod. Phys.} {\bf A8} (1993) 1823;
A. Ringwald, CERN preprint CERN-TH.6862 (1993), to appear in
{\it Proc. 4th Hellenic School on Elementary Particle Physics}, Corfu,
Greece, September 1992.
\bibitem{mo93a} D.A. Morris and A. Ringwald
in {\it Proc. 23rd Int. Cosmic Ray Conf.}, Calgary, July 1993,
ed. D.A. Leaky,
(U. of Calgary, 1993), Vol. 4, p. 407.
\bibitem{mo93b} D.A. Morris and A. Ringwald, CERN preprint
CERN-TH.6822 (1993), to appear in {\it Astropart.Phys.}.
\bibitem{mo91} D.A Morris and R. Rosenfeld, {\it Phys. Rev.}  {\bf D44}
(1991) 3530.
\bibitem{be92} L. Bergstr\"om, R. Liotta and H. Rubinstein,
{\it Phys. Lett.} {\bf  B276 } (1992) 231.
\bibitem{de92}
L. Dell'Agnello {\it et al.},
INFN Firenze preprint DFF 178/12 (1992),
to appear in {\it Proc. 2nd NESTOR Int. Workshop},
Pylos, October 1992.
\bibitem{kl84} N. Manton, {\it Phys. Rev. }{\bf D28} (1983) 2019;
Klinkhamer and N. Manton, {\it Phys. Rev. }{\bf D30} (1984) 2212.
\bibitem{za91} V. Zakharov, {\it Phys. Rev. Lett.} {\bf 67} (1991) 3650;
{\it Nucl. Phys.} {\bf B377} (1992) 501;
M. Maggiore and M. Shifman, {\it Nucl. Phys.}
 {\bf B371} (1992) 177; {\it ibid.} {\bf B380} (1992) 22;
G. Veneziano, {\it Mod. Phys. Lett.}
{\bf A7} (1992) 1661.
\bibitem{go92b} H. Goldberg, {\it Phys. Rev. Lett. }
{\bf 69} (1992) 3017.
\bibitem{ri91a} A. Ringwald and C. Wetterich, {\it Nucl. Phys.}
{\bf B353} (1991) 303; C. Wetterich, {\it Nucl. Phys. B } (Proc.
suppl.) {\bf 22A} (1991) 43.
\bibitem{fa90} G. Farrar and R. Meng, {\it Phys. Rev. Lett.}  {\bf 65}
(1990) 3377.
\bibitem{ri91b} A. Ringwald, F. Schrempp, and C. Wetterich,
{\it Nucl. Phys.} {\bf B365} (1991) 3.
\bibitem{vo80} L.V. Volkova, {\it Sov. J. Nucl. Phys.} {\bf 31} (1980) 784
[{\it Yad. Fiz.} {\bf 31} (1980) 1510.]
\bibitem{st91} F. Stecker, C. Done, M. Salamon, and P. Sommers,
{\it Phys. Rev. Lett.} {\bf 66} (1991)
2697; {\it ibid.}  {\bf 69} (1992) 2738 (Erratum);
in {\it Proc. Workshop on
High Energy Neutrino Astrophysics}, Honolulu, March 1992,
eds. V. Stenger, J. Learned, S. Pakvasa and X. Tata,
(World Scientific, Singapore, 1993), p. 1.
\bibitem{sz92}
A. Szabo and R. Protheroe,
in {\it Proc. Workshop on
High Energy Neutrino Astrophysics}, Honolulu, March 1992,
eds. V. Stenger, J. Learned, S. Pakvasa and X. Tata,
(World Scientific, Singapore, 1993), p. 24.
\bibitem{bi92}
P. Biermann,
in {\it Proc. Workshop on
High Energy Neutrino Astrophysics}, Honolulu, March 1992,
eds. V. Stenger, J. Learned, S. Pakvasa and X. Tata,
(World Scientific, Singapore, 1993), p. 86;
L. Nellen, K. Mannheim, and P. Biermann, {\it Phys. Rev.} {\bf D47}
(1993) 5270.
\bibitem{st94}
For a review see:
T. Stanev, in {\it these proceedings}.
\bibitem{gr66}
K. Greisen, {\it Phys. Rev. Lett.}  {\bf 16} (1966) 748;
G. Zatsepin and V.A. Kuzmin, {\it Pis'ma Zh. Eksp. Teor. Fiz.}
{\bf 4} (1966) 53 [{\it JETP Lett.} {\bf  4} (1966) 78].
\bibitem{be76}
V. Berezinskii and G. Zatsepin, in {\it Proc. 1976 DUMAND Workshop},
Honolulu, September 1976, ed. A. Roberts (Fermilab, Batavia, 1976) p. 15;
in {\it Proc. 15th Int. Cosmic Ray Conf.},
Plovdiv, Bulgaria, August 1977,
(Bulgarian Academy of Sciences, Sofia, 1977) p. 248;
V. Berezinsky and L. Ozernoy, {\it Astron. Astrophys.}  {\bf 98} (1981) 50.
\bibitem{st79} F. Stecker, {\it Astrophys. J.} {\bf 228} (1979) 919.
\bibitem{hi83} C. Hill and D. Schramm,
{\it Phys. Lett. } {\bf B131}  (1983) 247;
{\it Phys. Rev.} {\bf  D31} (1985) 564.
\bibitem{qu86} C. Quigg, M.H. Reno and T.P. Walker, {\it Phys. Rev}
{\bf D57} (1986) 774.
\bibitem{em92} B. Emerson, Ph.D. thesis, U. of Utah, 1992.
\bibitem{ba85} R. Baltrusaitis {\it et al.} (Fly's Eye Collaboration),
{\it Phys. Rev.} {\bf  D31} (1985) 2192.
\bibitem{ha84} For reviews see:
H. Harari, {\it Phys. Rep.} {\bf 104} (1984) 159;
W. Buchm\"uller, {\it Acta Physica Austriaca}, Suppl. {\bf XXVII}
(1985) 517.
\bibitem{mr92} S. Mrenna, {\it Phys. Rev.} {\bf D45} (1992) 2371.
\bibitem{do87} G. Domokos and S. Nussinov, {\it Phys. Lett.} {\bf B187}
(1987) 372; G. Domokos and S. Kovesi--Domokos, {\it Phys. Rev.}
{\bf D38} (1988) 2833.
\bibitem{ei83} E. Eichten, K. Lane and M. Peskin,
{\it Phys. Rev. Lett.} {\bf 50} (1983) 811.
\bibitem{pd92} K. Hikasa {\it et al.} (Particle Data Group),
{\it Phys. Rev.} {\bf D45} (1992) IX.12.
\bibitem{ar90} E. Argyres {\it et al.}, in {\it Proc. of the
Large Hadron Collider Workshop}, Aachen, Germany, October 1990,
eds. G. Jarlskog and D. Rein, (CERN 90--10, ECFA 90--133, Geneva, 1990),
Vol. II, p. 805.
\bibitem{ha86} F. Halzen, K. Hikasa and T. Stanev,
{\it Phys. Rev.} {\bf D34} (1986) 2061.
\bibitem{sc91} B. Schrempp, in {\it Proc. Workshop Physics at
HERA}, Hamburg, Germany, October 1991, eds. W. Buchm\"uller and
G. Ingelman, (DESY, Hamburg, 1992), Vol. 2, p. 1034.
\bibitem{sc91} P. Schleper, in {\it Proc. Workshop Physics at
HERA}, Hamburg, Germany, October 1991, eds. W. Buchm\"uller and
G. Ingelman, (DESY, Hamburg, 1992), Vol. 2, p. 1043.
\bibitem{bu87} W. Buchm\"uller, R. R\"uckl and D. Wyler,
{\it Phys. Lett.} {\bf B191} (1987) 442.
\bibitem{le93} M. Leurer, Weizmann Institute preprint,
WIS--93/90/Sept-PH, hep--ph/9309266, September 1993.
\bibitem{de93} M. Derrick {\it et al.} (ZEUS Collab.), {\it Phys. Lett.}
{\bf B306} (1993) 173; I. Abt {\it et al.} (H1 Collab.),
{\it Nucl. Phys.} {\bf B396} (1993) 3.
\bibitem{bu86} W. Buchm\"uller and D. Wyler, {\it Phys. Lett.}
{\bf B177} (1986) 377.
\bibitem{cdf92} S. Moulding (CDF Collab.), in {\it Proc. Seventh
Meeting of the APS}, Fermilab, USA, November 1992, ed. C.~Albright,
(World Scientific, Singapore, 1993) 1268.

\bibitem{containedshower}
J. Learned, private communication.

\end{thebibliography}
\end{document}